\documentclass[11pt]{article}
\usepackage[margin=1in]{geometry}
\usepackage{graphicx}
\usepackage{hyperref}

\def\beq{\begin{equation}}
\def\eeq#1{\label{#1}\end{equation}}
\def\eeqn{\end{equation}}

\def\beqa{\begin{eqnarray}}
\def\eeqa#1{\label{#1}\end{eqnarray}}
\def\eeqan{\end{eqnarray}}

\let\bar=\overbar

\def\Dslash{\not{\hbox{\kern-4pt $D$}}}
\def\dslash{\not{\hbox{\kern-2pt $\del$}}}

\def\msb{{\bar{\ssstyle M \kern -1pt S}}}

\def\Title#1{\begin{center} {\Large {\bf #1} } \end{center}}
\def\Author#1{\begin{center} {\normalsize {\sc #1} } \end{center}}
\def\Institution#1{\begin{center} {\normalsize {\it #1} } \end{center}}
\def\Abstract#1{\noindent {\normalsize {\bf Abstract:} {\normalfont #1}}}
\def\Conference{\vspace{4mm}\begin{raggedright} {\normalsize {\it Talk presented at the 2019 Meeting of the Division of Particles and Fields of the American Physical Society (DPF2019), July 29--August 2, 2019, Northeastern University, Boston, C1907293.} } \end{raggedright}\vspace{4mm}}

\begin{document}

%
%

\Title{Exploration of the Dark Sector with the Fermilab Dimuon Experiment}

\Author{Vassili Papavassiliou}

\Institution{Physics Department\\ New Mexico State University, Las Cruces, NM, USA\\
for the SpinQuest/E-1039 Collaboration}

\Abstract{Searches for dark-matter particles at the GeV mass scale has been
receiving much attention in the last several years, partly motivated
by the failure of direct and indirect searches of heavier candidates
to produce a signal. The SpinQuest dimuon experiment in the 120-GeV
Main-Injector proton beam at Fermilab, currently in the commissioning
stage, is uniquely equipped to search for dark photons and dark Higgs
particles produced in a 5-m long iron beam dump with masses in the
range 0.2 - 10 GeV, running in a parasitic mode. This only requires
a modest upgrade of a displaced-vertex trigger with acceptance for
dark-sector particles decaying into dimuons inside or downstream of
the dump. We discuss the physics reach of such a run, the status,
and some additional future prospects.}

\Conference

%
%

\section{Introduction}
Over the last four decades, experimental and theoretical efforts for
understanding the nature of dark matter have been dominated by the
WIMP: Weakly Interacting Massive Particles at the weak-interaction
mass scale, $\mathcal{O}(100\mathrm{~GeV})$. Such particles arise
naturally in supersymmetric extensions of the standard model of particle
physics and with the right relic abundance to explain the observed
density; this is the so-called ``WIMP miracle.'' The situation started
to change during the last few years, however, due to the failure to
observe any such particles in direct and indirect searches: production
in colliders, especially the LHC; scattering from ordinary matter in
detectors; and indirectly in searches for high-energy neutrinos or photons
from dark-matter particle collisions and annihilations, for example in the
center of the galaxy. This has created opportunities for alternative models
of dark matter to receive increased attention; for a recent summary of the
situation see, for example, Ref.~\cite{Bertone:2018xtm}.

There is no shortage of alternative DM models, from extremely light
bosons, with de Broglie wavelengths of the order of 1~kpc (``fuzzy dark
matter''), to massive primordial black holes. One particular class of
models predicts the existence of a hidden, ``dark sector'' which makes
its presence revealed almost exclusively through the gravitational
interaction. However, in some models ``portal interactions'' can couple
the dark sector to the Standard Model particles; for example, an
additional U(1)$_X$ symmetry can result in a dark photon $A^\prime$
which can mix with the SM U(1)$_Y$ photon with coupling $\epsilon$,
described by an additional term in the Lagrangian of the form
\begin{equation}
  \mathcal{L}_\mathrm{mix}=\frac{\epsilon}{2}F^{Y,\mu\nu}F^X_{\mu\nu}
\end{equation}
(vector portal); the dimensionless, kinetic-mixing parameter $\epsilon$
and the mass of the dark photon $m_{A^\prime}$ determine the strength of the
interaction with ordinary matter. Similarly, a dark Higgs $\phi$ can
mix with the SM Higgs $H$ through a term
\begin{equation}
  \mathcal{L}_\mathrm{mix}=\mu\phi|H^\dagger H|
\end{equation}
(scalar portal); in this case the interaction is determined by the mass
of the dark Higgs, $m_\phi$, and the mixing angle $\theta$, related to
$\mu$, the electroweak scale $\nu$, and the mass of the SM Higgs $m_h$
by $\theta=\mu\nu/m_h^2$. In turn, the interaction strenghts determine
both the production cross sections from ordinary matter and the decays
into SM particles.

While the masses of these particles can vary over more than 30 orders
of magnitude in the different models, recent hints of anomalies, in the
muon $g-2$ value and positron excess in cosmic rays, have directed much
interest in the region from $\mathcal{O}(1\mathrm{~MeV})$ to
$\mathcal{O}(10\mathrm{~GeV})$
for the dark photon mass. Much of this region can be probed efficiently
in high-luminosity, fixed-target experiments and many are under way or
being considered. In the following, we describe the reach of the Fermilab
dimuon experiment SpinQuest/E-1039, running in a parasitic mode where
events from the beam dump are considered. Additional details can be found
in Ref.~\cite{Liu:2017ryd}.

\section{The Experiment}
\begin{figure}[htb]
 \centering
 \includegraphics[height=2.25in]{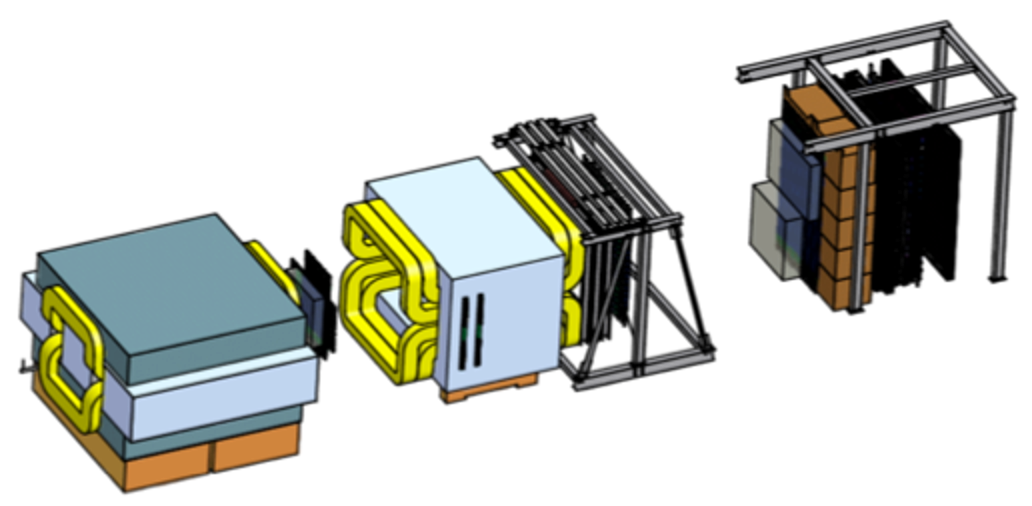}
 \caption{The SpinQuest spectrometer. The beam enters from the lower left.
   The target, located upstream, is not shown. Magnet coils for the FMag and
   KMag dipole magnets are shown in yellow. The beam dump, 5~m of iron, is
   located inside the first dipole, FMag. Both magnets bend in the horizontal
   plane. Tracking chambers and scintillator hodoscopes are located both
   upstream and downstream of KMag. At the most downstream part of the
   spectrometer there is a muon identifier consisting of a hadron absorber
   followed by scintillator hodoscopes and  proportional tubes.}
 \label{fig:SpinQuest}
\end{figure}

SpinQuest/E-1039 is a dimuon experiment in the Fermilab 120-GeV,
Main-Injector, proton beam. It is a follow-up, with improvements, of
the SeaQuest/E-906 experiment which studied the composition of the
nucleon sea through the Drell-Yan process. SpinQuest uses a polarized,
frozen-ammonia (NH$_3$) or deuterated ammonia (ND$_3$) target to study
QCD effects due to transversely-polarized partons in the nucleon through
transverse, single-spin asymmetries. The thin targets, approximately
6\% of an interaction length, combined with the location of the 5-m
thick iron beam
dump and hadron absorber upstream of the spectrometer, present an
opportunity to search, in a parasitic mode, for rare, long-lived
particles produced in the beam dump. These can propagate without
intracting until they decay, either in the dump or immediately downstream,
into pairs of muons, which then enter the
spectrometer. Figure~\ref{fig:SpinQuest} shows the SpinQuest spectrometer.

A dedicated trigger was designed to complement the main physics trigger,
which only selects events from the target, upstream from the dump. The
dark-photon/dark-Higgs trigger is based on a new set of fine-grained,
scintillating-strip hodoscopes that measure the $Y$-coordinate (non-bending
view) of the two muons and will determine the $Z$-position (along the beam
direction) of the vertex. In addition, improvements to the data-acquisition
system can accommodate event rates an order of magnitude higher than at
SeaQuest.

\begin{figure}[htb]
 \centering
 \includegraphics[height=2.25in]{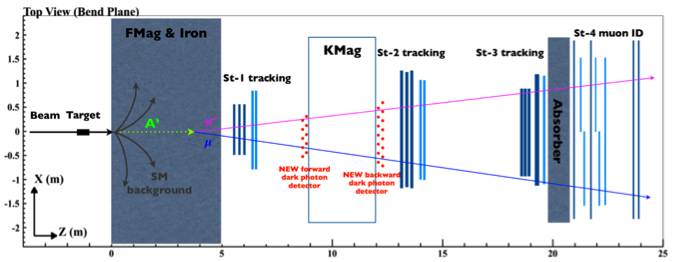}
 \caption{Dark photon production and decay in the SpinQuest experiment.
   Protons from the beam that do not interact in the
   $\mathcal{O}(6\%\lambda_I)$
   target will interact in the first few centimeters of the iron beam
   dump. Any weakly-interacting, long-lived particles produced can be
   detected through a possible dimuon decay mode. The new, fine-grained
   trigger hodoscopes (red) will serve for triggering on such displaced
   vertices.}
 \label{fig:DPDecay}
\end{figure}

Figure~\ref{fig:DPDecay} shows the production and decay of a dark photon
in the beam dump, inside the focusing magnet FMag. The new trigger
hodoscopes, upstream and downstream of the magnet repurposed from the KTeV
experiment (KMag) are shown in red.

\section{Sensitivity}
\begin{figure}[htb]
 \centering
 \includegraphics[height=5in]{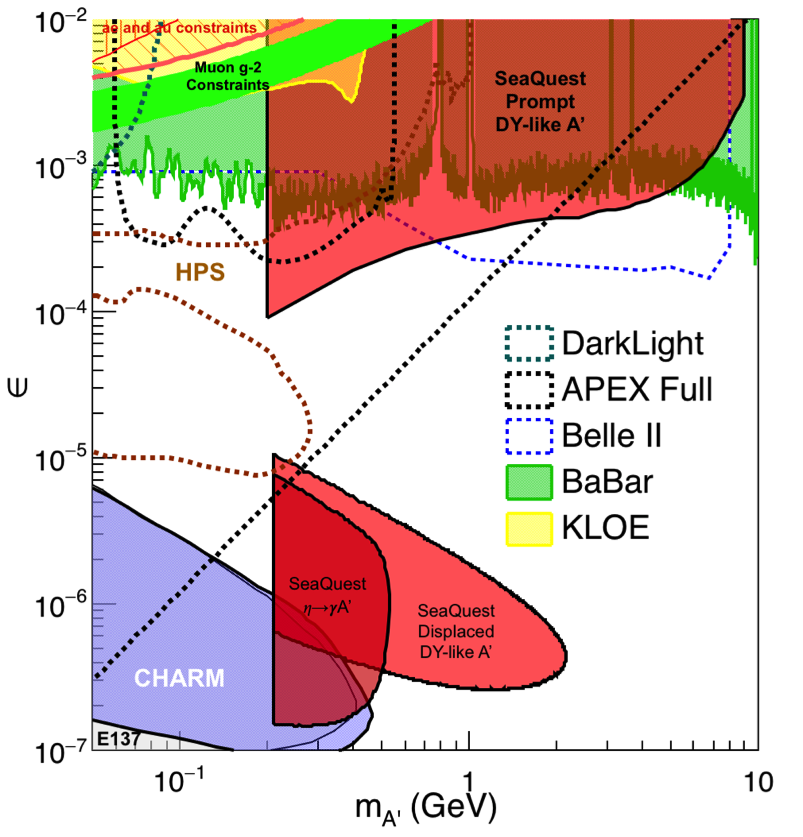}
 \caption{Reach of current and projected future experiments on the search
   for dark photons (red areas). Expected gaps due to the $\phi$, $J/\psi$,
   and $\psi^\prime$ resonances will depend on the experimental resolution
   and are not yet shown. From Ref.~\cite{Liu:2017ryd}.}
 \label{fig:DPSensitivity}
\end{figure}

The SpinQuest experiment is approved for $1.4\times10^{18}$ protons on
target (POT) over a period of two years. The parasitic mode will have
sensitivity to dark-photon masses
from about 0.2~GeV (the dimuon threshold) to almost 10~GeV, with gaps at
the masses of the $\phi$ and $\psi$ resonances. In this range, production
mechanisms are $\pi$, $\eta$, $\omega$, and $\eta^\prime$ decay, proton
bremsstrahlung, and the Drell-Yan process; where in all cases a real or
virtual SM photon mixes with the dark photon (for example
$\pi^0\rightarrow\gamma A^\prime$). Figure~\ref{fig:DPSensitivity} shows
the reach in the $\epsilon$ vs.~$m_{A^\prime}$ plane for a dark photon with
mass $m_{A^\prime}$ and mixing parameter $m_{A^\prime}$, compared with other,
current and planned experiments.

The expected sensitivity (95\% exclusion region) from
a parasitic run during the SpinQuest/E-1039 (previously SeaQuest/E-906)
experiment with $1.4\times10^{18}$ protons on target
is shown as red areas; the area at the top of the figure (corresponding
to larger values of $\epsilon$) results from prompt, Drell-Yan-like
$A^\prime$, while the areas at the bottom (small $\epsilon$) arises from
either the decay $\eta\rightarrow\gamma A^\prime$ or displaced, DY-like
$A^\prime$. Exclusion zones from past experiments, BaBar (green),
KLOE (yellow), and CHARM (magenta) are also shown. Expected capabilities
of other planned experiments are displayed with dashed lines. Searches
at the LHC are mostly sensitive to dark photons with masses larger than
10~GeV and are not shown on this figure. The straight, diagonal, black,
dashed line represents a relationship $m_{A^\prime}\approx\sqrt{\epsilon}m_Z$
preferred in some theoretical models. A darker green area on the top
left corner represents the constraints from the muon $g-2$ anomaly if
this model is invoked to provide the explanation. 

\begin{figure}[htb]
 \centering
 \includegraphics[height=5in]{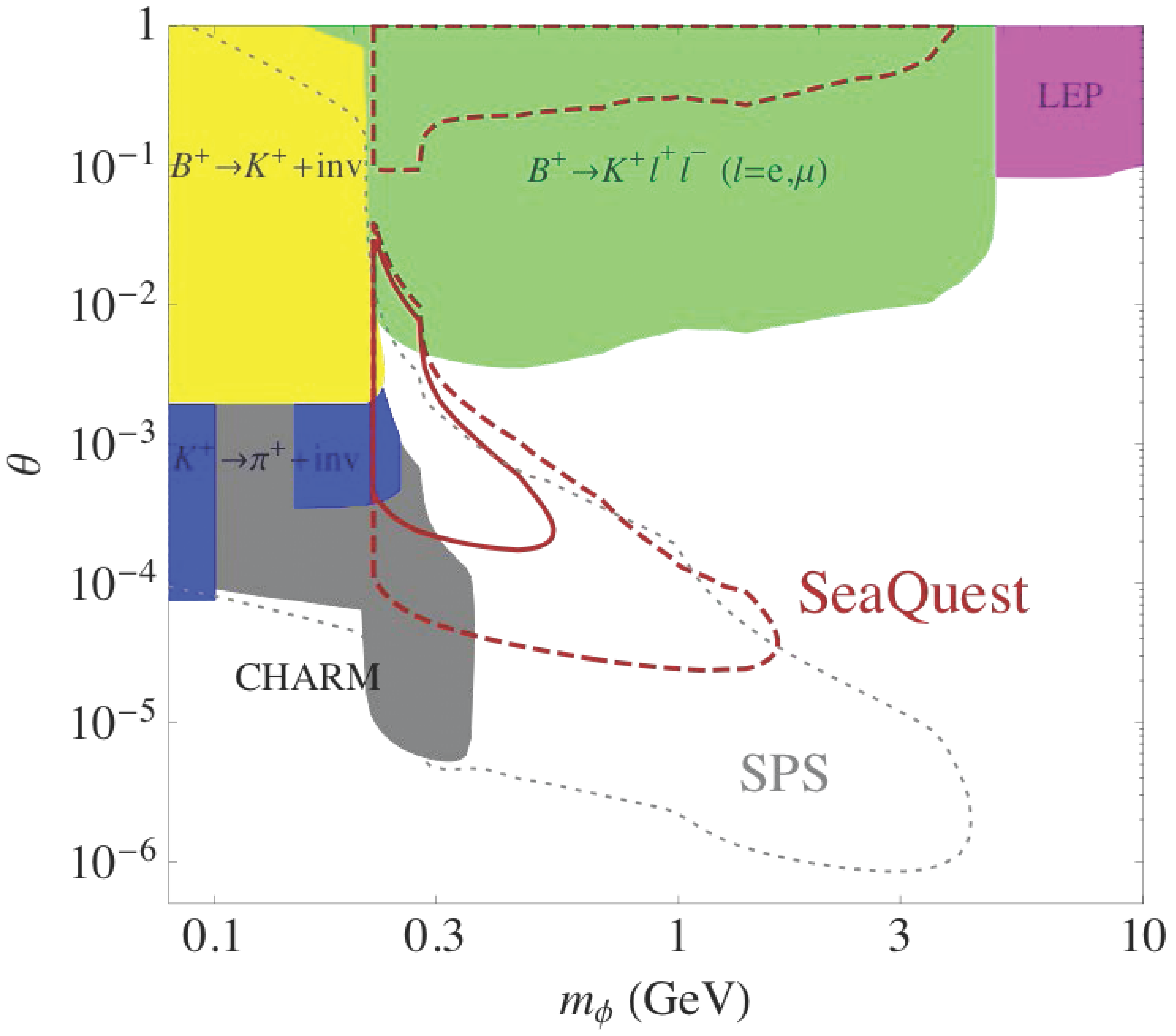}
 \caption{Projected sensitivity for a dark-Higgs search in the SpinQuest
   experiment (solid red line) and a proposed, much higher luminosity,
   dedicated run to follow (dashed red line). A main competitor, the SHiP
   experiment (dotted line; ``SPS'') is projected to start taking data
   in 2026. Solid colored areas show current exclusion limits. From
   Ref.~\cite{Liu:2017ryd}.}
 \label{fig:DHSensitivity}
\end{figure}

Similarly, Fig.~\ref{fig:DHSensitivity} shows the reach of a search
for a dark Higgs particle in the $m_\phi$-$\theta$ plane. Present
constraints, mainly from $B$ and $K$ meson decays, are shown as filled
areas. The solid, red line shows the 95\% sensitivity from the parasitic
run with $1.4\times10^{18}$ POT, while the dashed, red line represents the
anticipated sensitivity from a dedicated, future run with $4\times10^{20}$
POT, briefly described later. The sensitivity overlaps much with that of
the proposed SHiP (Search for Hidden Particles) experiment at the CERN
SPS (dotted line, labelled SPS) which however will not start taking data
until after the third long shutdown of the LHC, ca.~2026, providing a
window of opportunity for the current and the proposed, dedicated run to
make an impact on this channel as well.

\section{Status}
The parasitic run was approved by the Fermilab Program Advisory Committee
and will take place during the SpinQuest run starting in late 2019.
As of this writing, the experiment is in the commissioning stage and
undergoing safety reviews. Nightly
data-acquisition test runs are performed and the trigger, online monitoring,
and event displays are operational. The polarized target will be installed
within the upcoming months. Beam is expected before the end of 2019. A new
letter of intent has been submitted to the PAC for a run that will have
much higher sensitivity, as described in the following section.

\section{Future Prospects}
The collaboration is proposing a dedicated run, at the end of the
polarized-target run, with much higher integrated luminosity of up to
$4\times10^{20}$ POT. The lower mass limit can also be pushed down to
almost 1~MeV by detecting in addition the dielectron decay mode. For that,
an electromagnetic calorimeter (EMCal) would be installed, transferred from
the recently completed PHENIX experiment at RHIC. Dielectron decays are
of course only detected if they occur downstream of the iron absorber,
limiting the reach via this mode.

\begin{figure}[htb]
 \centering
 \includegraphics[height=2in]{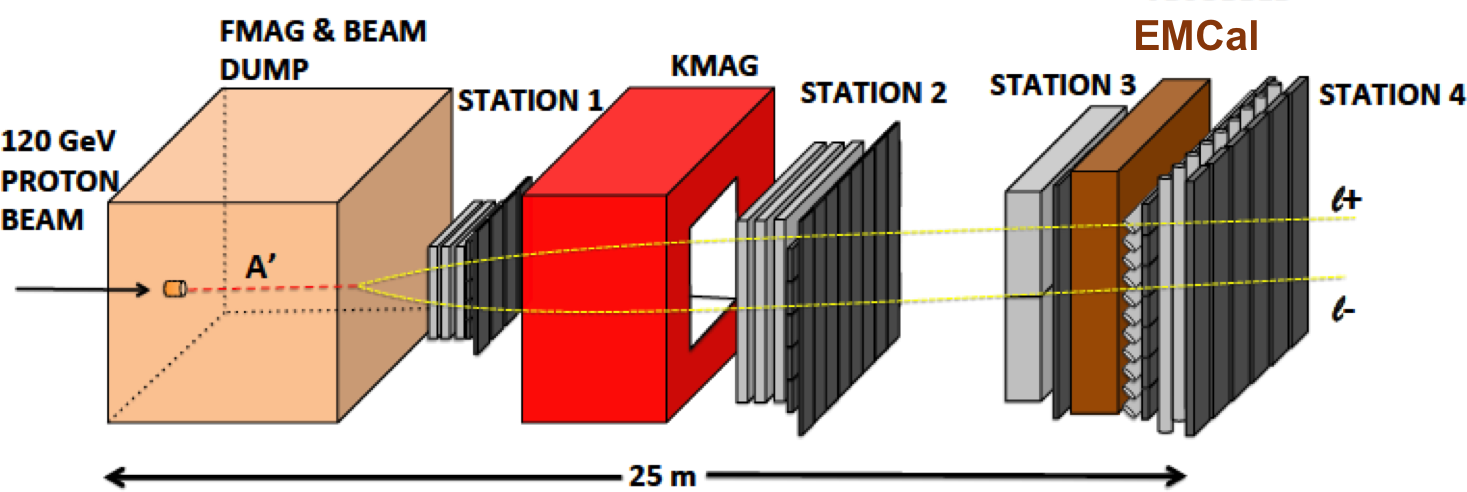}
 \caption{Upgraded detector for a dedicated run, including dielectron
   detection capabilities with the addition of an electromagnetic calorimeter.}
 \label{fig:Upgrade}
\end{figure}

Figure~\ref{fig:Upgrade} shows the proposed, upgraded SpinQuest
spectrometer with the EMCal installed between the two most downstream
detector stations. Additional projections under various assumptions
regarding the capabilities of such an experiment and comparisons with
other proposed experiments can be found in Ref.~\cite{Berlin:2018pwi}.

\section*{Acknowledgements}
The author thanks the SpinQuest/E-1039 collaboration for helpful
comments and discussions during the preparation of this manuscript.
This work was partly supported by a grant to NMSU by the US Dept.~of
Energy, Office of Science.

\end{document}